\documentclass[11pt]{article}
\usepackage{amssymb}
\usepackage{amsmath}
\usepackage{graphicx}
\usepackage{color}

\def\beq{\begin{eqnarray}}    
\def\eeq{\end{eqnarray}}      

\def\beq{\begin{eqnarray}}    
\def\eeq{\end{eqnarray}}      

\newcommand{\be}{\begin{equation}}
\newcommand{\ee}{\end{equation}}

\begin{document}



 \hyphenation{cos-mo-lo-gi-cal
sig-ni-fi-cant par-ti-cu-lar}




\begin{center}
{\bf \Large
\textit{Fundamental Constants in Physics and Their Time Variation}}\footnote{Preprint of the Preface to the Special Issue on \textit{Fundamental Constants in Physics and Their Time Variation}, published in Modern Physics Letters A
Vol. 30, No. 22 (2015) 1502004 [5 pages]
[DOI: 10.1142/S0217732315020046]  Guest Editor of the Special Issue: J. Sol\`a. }

\vspace{0.5cm}
({\large \bf Preface})
\end{center}
\vspace{1cm}
\begin{center}

{\bf \large Joan Sol\`a}\\

Universitat de Barcelona\\

joan.sola@ub.edu

\end{center}

\vspace{0.5cm}

In this Special Issue, we present a
collection of fresh new articles, together with some review works, written by first-rate researchers in the field that offer the state-of-the-art on the subject of \textit{Fundamental Constants in Physics and Their Time
Variation}. There is no doubt that this is one of the hottest
subjects in modern theoretical and experimental physics, with
potential implications in all fundamental areas of physics research,
such as particle physics, gravitation, astrophysics and cosmology.

\vspace{0.5cm}
The history of this subject traces back mainly to Dirac's pioneering
work in the thirties on the ``large number hypothesis''\,\cite{Dirac} , from which
a time evolution of the gravitational constant $G$ was suggested, and later on the first discussions on new forms of the principle of equivalence emerged\,\cite{Dicke} and finally triggered the Jordan, Fierz and Brans-Dicke approach to gravity\,\cite{JFBD}, in which General Relativity was extended to accommodate variations in $G$. It
also triggered subsequent speculations by Gamow\,\cite{Gamow} and others on the
possible variation of the fine structure constant.

\vspace{0.5cm}
Despite the
initial difficulties, these seminal works were a real spur to start
changing the mentality and the strong prejudice on the supposedly
imperturbable and ``sacrosanct'' rigid status of the constants of Nature. It is amazing
to realize nowadays how much we have opened our minds to the new horizons that these ideas offered since those early times. Modern investigations on this subject are performed
not only at the theoretical but also at the experimental level, both
in the lab (through high precision quantum optic techniques) and in
the astrophysical domain (using absorption systems in the spectra of
distant quasars). In the last decade and a half different
astrophysical observations of this kind have suggested positive
evidence on the time change of the fine structure constant,
$\alpha_{\rm em}$, and there have also appeared intriguing
indications of a possible space (local) variation of the same
quantity, cf.\,\cite{UzanChiba,CalmetKeller,Webb1999,Webb2011,Stadnik2015}. Measurements of these two types of effects have been
claimed in the literature at the level of  $4-5\sigma$, but they need
independent confirmation by other groups. Similarly, the
dimensionless ratio $\mu\equiv m_p/m_e$ (of proton to electron
masses) has been carefully monitored (once more using quasar absorption lines) with the result that significant time variation of $\dot{\mu}/\mu$ at a $\sim4\sigma$
level has been reported, although still unconfirmed by other observations.
Future high precision quantum optic experiments in the lab are also
planned to test the possible time variation of these observables, and they will most likely be competitive.

\vspace{0.5cm}
Quite obviously this is a very active field of research. Exciting new results are expected soon which could  significantly modify our current scientific paradigms. If we attend
to what we know about the energy budget of the Universe, which is
believed to consists roughly of only 5\% of baryonic matter (atoms),
25\% of dark matter (DM) and 70\% of dark energy (DE), it is pretty
obvious that our knowledge on its composition is more than limited
and hence leaves much to be desired. It is not surprising that many
researchers, spurred by the positive observational hints and the
unsatisfactory theoretical situation, have seriously adhered to the
possibility that the so-called fundamental ``constants'' of nature
can be, in reality, slowly varying quantities possibly related to
underlying fundamental theories of the elementary interactions. These theories are unknown at present, but they might help explaining the origin of the hidden components of the DM and the nature of the DE, which dominate by far the structure and fate of our Universe.

\vspace{0.5cm}
If we descend a further level of theoretical detail and scrutinize
the status of the best model we have to date for studying the subatomic word, namely the standard model (SM) of the strong and electroweak (EW)
interactions, we find that it consists of many parameters whose ultimate
origin and interrelationship remains completely unknown. For
example, we can identify 27 (presumably) independent fundamental
constants, to wit: the QED fine structure constant $\alpha_{\rm
em}=e^2/4\pi$, the $SU(2)_L$ gauge coupling $g$ of the EW
interactions, the gauge coupling constant of the strong interactions
$g_s$, the masses $M_{W,Z}$ of the weak gauge bosons, the mass
$M_{H}$ of the Higgs boson (currently a measured parameter), the 12
masses of the quarks and leptons, the 3 mixing angles of the quark
mass matrix, a CP-violating phase, the $3$ mixing angles in the
lepton sector, a CP-violating phase and two additional phases, if
the neutrino masses are Majorana masses.
We do not have at present an explanation for the large variety of couplings, masses and mixing
angles in the SM. Admittedly the model works quite well since
all the other observables (e.g. cross-sections, decay rates etc) can
be explained satisfactorily with those 27 parameters, and in many cases
not only at the tree level but also at high loop order.

\vspace{0.5cm}
Recently the
LHC collaborations at CERN have tagged a scalar particle, with a mass of around $125$ GeV\,\cite{HiggsDiscovery}. The new particle
at the moment seems to carry all the physical attributes for being
identified with the Higgs boson (or at least ``a'' Higgs boson
within some popular extension of the SM containing several such
spinless particles). In that case the entire particle content of the
SM, and eventually of the most favorite extensions thereof, will have been elucidated.
However, we do not have at present a real, truly profound, understanding of the theoretical
structure of the SM. Some very obvious questions remain still unanswered, let us just mention three of utmost importance: i) why the number
of parameters is so large? ii) why just three families of quarks and leptons? iii) why do we have such a large hierarchy of fermion masses ranging from extremely light neutrinos to the super heavy top quark?  Ultimately our crucial question boils down to this one: are all these parameters really fundamental
constants of nature? Somehow we expect that there should be a deeper
correlation among them, which might eventually become apparent in the
context of a more complete theory.

\vspace{0.5cm}
If, in addition, we include the Einstein-Hilbert action in the above
field theoretical structure, two more fundamental ``constants''
enter into play, both of them dimensionful; namely, Newton's
gravitational coupling $G$ and the cosmological constant $\Lambda$
(or ``CC term''). The gravity constant can be related to a mass
parameter through $G=\hbar c/M_P^2$, where $M_P\simeq 1.221\times
10^{19}$ GeV defines the Planck mass, which is the largest mass
scale in the Universe. However, the relation of that mass with the
gravitational strength $G$ depends on the Planck constant $\hbar$
and the velocity of light in vacuo, $c$. This implies that (relativistic)
quantum theory is inherently involved in that link, and is of course the
reason why it is usually stated that, at energies above $M_P c^2$,
quantum gravity should start playing a crucial role in the
description of the Universe.

\vspace{0.5cm}
As for the cosmological constant $\Lambda$, it is perhaps the most
mysterious of all the fundamental ``constants'' of the Universe\,\cite{Weinberg}. If
we combine $G$, $\Lambda$ and $c$, we may form the quantity
$\rho_{\Lambda}\,c^2=\Lambda c^4/(8\pi G)$ which has dimension of
energy density. It is usually called the \textit{vacuum energy
density}, and it is the standard candidate for explaining the DE.
The so-called $\Lambda$CDM model (or concordance model of cosmology)
is precisely based on this assumption. Notice that $\rho_{\Lambda}$
has dimension of mass density and its  (observationally measured)
value is of order $\rho_{\Lambda}\simeq 10^{-29} g/cm^3$. Obviously it is
a very small density parameter, just around $0.7$ times the
value of the critical density $\rho_c$ of the Universe. In natural
units ($\hbar=c=1$) we can rewrite it as $\rho_{\Lambda}\sim 10^{-11}$
eV$^4$. It follows that the mass scale associated to such density, i.e. the quantity
$m_{\Lambda}$ that satisfies $\rho_{\Lambda}\sim m_{\Lambda}^4$, is
of order of a millielectrovolt: $m_{\Lambda}\sim 10^{-3}$ eV. This
is quite astonishing if we take into account that all of the mass
scales in the SM (up to perhaps the mass of a very light neutrino)
are exceedingly much heavier than $m_{\Lambda}$. For instance,
$m_e/m_{\Lambda}\sim 5\times 10^{8}$ for the electron and
$M_W/m_{\Lambda}\sim  10^{14}$ for a weak gauge boson. Thus we find
that the electroweak vacuum energy density is typically
$M_W^4/m_{\Lambda}^4\sim 10^{56}$ times bigger than the cosmic
vacuum energy density! This preposterous result  is at the root of the
so-called \textit{cosmological constant problem}, perhaps the
biggest conundrum of theoretical physics ever\,\cite{Weinberg} -- see also \,\cite{JSP2013} for a vivid account, and
\cite{JSPEssay2015} for possible implications on the physics of the early Universe. It rises one more crucial question (the fourth one, to add to the previous list): what is the ultimate nature of the quantum vacuum; and, more specifically, how to eventually reconcile the vacuum of the SM of strong and electroweak interactions with the observed value of the cosmological constant?

\vspace{0.5cm}
Remarkably, two years from now, in 2017, it will be the centenary of the
introduction of the CC term by Einstein in his gravitational field
equations\,\cite{Einstein1917}, and will also be half century of the history of the cosmological constant problem as such\,\cite{Zeldovich67}, namely the (as yet)  impossible
task of harmonizing the quantum vacuum with the cosmic vacuum.
Since then we have been able to
measure the value of $\Lambda$ with high precision and by different and independent
astrophysical and cosmological observations\,\cite{Planck}, but we still lack a truly profound theoretical
understanding of its real meaning, and we fully ignore the reason for
the tremendous, truly devastating, mismatch between the observed and the  predicted
value of the CC in quantum field theory or string theory.

\vspace{0.5cm}
From the above considerations it should be clear that the subject of
the fundamental constants and their possible variation (both in time
and perhaps also in space) is at the root of some of the most
fundamental problems of physics.
In this volume seven works are
presented which address different aspects of the fundamental
``constants'' of nature and their possible time and/or space
variation:

\vspace{0.5cm}
\begin{itemize}

\item The article by  Xavier Calmet and Matthias Keller\,\cite{CalmetKeller}
describes the state of the art, both at the theoretical and
experimental level, of the fundamental constants and their possible
cosmological evolution;

\item The work by John D. Barrow and Jo\~{a}o
Magueijo\,\cite{BarrowMagueijo} emphasizes the possibility of a local variation of the fine
structure constant and proposes a concrete theoretical framework to
account for it, namely one where the vacuum is regarded
as a dielectric medium with unusual properties;

\item The next work, authored by Bennie F.L. Ward\,\cite{Ward},
performs an attempt at estimating the numerical value of the
cosmological constant on the basis of the resummation techniques in
quantum General Relativity and the idea of running $\Lambda$ and
$G$;

\item Related in part to the previous one, Spyros Basilakos\,\cite{Basilakos} reviews in his contribution the features of the cosmic expansion and structure
formation in the context of the running vacuum cosmologies;

\item The article by Salvatore Capozziello and Gaetano Lambiase\,\cite{CapozzieloLambiase} plunges into
the physics of variable $G$ models by considering the propagation of
quantum particles in the framework of spherically symmetric
solutions of  Brans-Dicke spacetime, and evaluates the detection of
these effects using gamma ray bursts;

\item Some aspects of the tough cosmological constant
problem and its connection with the notion of vacuum energy are
addressed in the review contribution by Steven Bass\,\cite{Bass}; in particular the possibility that the cosmological
constant plus LHC results might hint at critical phenomena near the Planck scale;

\item Finally, Harald
Fritzsch and the author\,\cite{FritzschSola} explore the possibility that there is a consistent
feedback between the micro and macro cosmos, namely a subtle crosstalk that could explain the time variation of the fundamental constants of the subatomic world (masses and
couplings) in correlation with the time evolution of the
cosmological term (appearing as dynamical dark energy) and Newton's coupling.

\end{itemize}

\newpage
We are confident that the set of articles contained in this
Special Issue of Modern Physics Letters A will provide the readers
with an up-to-date precise of the theoretical and experimental
situation of the fundamental ``constants'' and the exciting possibility that they may have undergone a non-negligible evolution throughout the cosmic history. We also hope they can serve as an stimulus for future work on this field, and especially for the development of the necessary new ideas that will contribute in the future to our better understanding of some of the most puzzling problems of gravitation, particle physics and cosmology.

\vspace{0.5cm}



\begin{thebibliography}{99}

\bibitem{Dirac}
P. A. M. Dirac, {Nature}\, {\bf 139} (1937) 323;
Proc. Roy. Soc. London A {\bf 165} (1938) 198.

\bibitem{Dicke} R.H. Dicke, Rev. Mod. Phys. {\bf 29}  (1957) 355; Nature {\bf 192} (1961) 440.

\bibitem{JFBD} P. Jordan, \textit{Schwerkraft und Weltall. Grundlagen der theoretischen Kosmologie} (Friedr. Vieweg \& Sohn, Braunschweig, 1955);
    M. Fierz, Helv. Phys. Acta {\bf 29}  (1956) 128;
    C. Brans and R. H. Dicke, Phys. Rev. {\bf 124} (1961) 925;
    R. H. Dicke, Phys. Rev. {\bf 125} (1962) 2163.

\bibitem{Gamow} G. Gamow, Phys. Rev. Lett. {\bf 19}  (1967) 759.


\bibitem{UzanChiba} J-P. Uzan,
Liv. Rev. Rel. {\bf 14} (2011) 2;
T. Chiba, Prog. Theor. Phys. 126 (2011) 993.

\bibitem{CalmetKeller}  X. Calmet and M. Keller, \textit{ 	
Cosmological Evolution of Fundamental Constants: From Theory to Experiment}, Mod. Phys. Lett. A{\bf 30} (2015) 1540028 [13 pages].

\bibitem{Webb1999} J. K. Webb, V. V. Flambaum, C. W. Churchill, M. J. Drinkwater and J. D. Barrow, Phys. Rev. Lett. {\bf 82} (1999) 884.

\bibitem{Webb2011} J. K. Webb et al., Phys. Rev. Lett. {\bf 107} (2011) 191101; J. A. King et al., Mon. Not. R. Astron. Soc. {\bf 422} (2012) 3370.

\bibitem{Stadnik2015} Y. V. Stadnik and V. V. Flambaum, Phys. Rev. Lett. {\bf 114}  (2015) 161301; e-Print: arXiv: 1503.08540.

\bibitem{HiggsDiscovery} The ATLAS Collab., Phys. Lett. B{\bf 716} (2012) 1; The CMS Collab., \textit{ibid}. B{\bf 716} (2012) 30.

\bibitem{Weinberg}S. Weinberg,
Rev. Mod. Phys. {\bf 61} (1989) 1;
V. Sahni, A. Starobinsky, Int. J. of Mod. Phys. A{\bf 9} {(2000)} {373};
T. Padmanabhan, Phys. Rept.  {\bf 380} (2003)  235;
P.J. Peebles, B. Ratra, Rev. Mod. Phys. {\bf 75}
    (2003) 559.



\bibitem{JSP2013} J. Sol\`a,  \textit{Cosmological constant and vacuum energy: old and new ideas}, J. Phys. Conf. Ser. {\bf 453}  (2013) 012015; J. Sol\`a, and A. G\'omez-Valent,
    Int. J. of Mod. Phys. D{\bf 24} (2015) 1541003.

\bibitem{JSPEssay2015}  J. Sol\`a,  \textit{ 	
The cosmological constant and entropy problems: mysteries of the present with profound roots in the past}, e-Print: arXiv:1505.05863  (to appear in Int. J. Mod. Phys. D.).


\bibitem{Einstein1917} A. Einstein, \textit{Kosmologische Betrachtungen zur allgemeinen
Relativit\"atstheorie}, {Sitzungsberichte der K\"oniglich
Preu{\ss}ischen Akademie der Wissenschaften zu Berlin},
phys.-math. Klasse VI (1917) 142-152.

\bibitem{Zeldovich67} {Y. B. Zeldovich}, \textit{ 	
Cosmological constant and elementary particles}, JETP Lett. {\bf 6} (1967) 316, Pisma Zh.Eksp.Teor.Fiz. {\bf 6} (1967) 883; \textit{Cosmological constant and the theory of elementary particles},  Sov. Phys. Usp. {\bf 11}  (1968) 381.


\bibitem{Planck}  	
Planck 2013 results. XVI (P.~A.~R.~Ade {\it et al.}). \textit{Cosmological parameters},
   Astron. Astrophys. {\bf 571} (2014) A16;  	
Planck 2015 results. XIII. \textit{Cosmological parameters}
e-Print: arXiv:1502.01589.

\bibitem{BarrowMagueijo} J. D. Barrow, J. Magueijo,
\textit{Local Varying-Alpha Theories}, Mod. Phys. Lett. A{\bf 30} (2015) 1540029 [16 pages].

\bibitem{Ward} B.F.L. Ward, \textit{Running of the cosmological constant and estimate of its value in quantum general relativity},  Mod. Phys. Lett. A{\bf 30}  (2015) 1540030 [15 pages].

\bibitem{Basilakos} S. Basilakos, \textit{Cosmic expansion and structure formation in running vacuum cosmologies}, Mod. Phys. Lett. A{\bf 30} (2015) 1540031 [17 pages].

\bibitem{CapozzieloLambiase} S. Capozziello, G. Lambiase, \textit{Propagation of quantum particles in Brans–Dicke spacetime: The case of gamma ray bursts}, Mod. Phys. Lett. A{\bf 30}  (2015) 1540032 [13 pages].


\bibitem{Bass} S. D. Bass, \textit{Vacuum energy and the cosmological constant}, Mod. Phys. Lett. A{\bf 30} (2015) 1540033  [15 pages].

\bibitem{FritzschSola} H. Fritzsch, J. Sol\`a, \textit{Fundamental constants and cosmic vacuum: The micro and macro connection},  Mod. Phys. Lett. A{\bf 30} (2015) 1540034 [16 pages].





\end{thebibliography}
\end{document}